\documentclass[12pt,reqno]{article}
\usepackage{amsmath}
\usepackage{graphicx}
\usepackage{color}
\begin{document}

\title
\centerline{\bf\large Self Magnetic Field and Current-loop of Electron with Five Different Radii and Intrinsic Properties}

\vspace{12pt}

\centerline{{\bf S. Ghosh}$^{\rm *}$, {\bf M. R. Devi}, {\bf A. Choudhury} and  {\bf J. K. Sarma }}
\centerline{ HEP Laboratory, Department of Physics, Tezpur University, Tezpur, Assam, 784 028, India} 
\centerline{{\it gsovan@gmail.com}$^{\rm *}$}

\bigskip
\bigskip

\begin{abstract}
\leftskip1.0cm
\rightskip1.0cm

With the rotation of charge in a circular path current-loop is generated. This results in the production of magnetic fields in a charged body. This self (internal) magnetic field is treated here with five different radii of electron.  Closed loop line integral of magnetic induction vector or the integral form of the Ampere's circuital law is expressed in terms intrinsic properties of electron. 
\end{abstract}

\section{Introduction}
\paragraph \ 
Electromagnetic theory puts electric and magnetic fields on the same footing to explain different physical phenomena[1]. The behaviour of charged particles can also be explored by probing its behaviour in uniform[2] and non-uniform[3] magnetic field. As well as the measures of magnetic field also come out from the behavior of charged particles and their properties.

\ The oldest discovery of particle physics is electron, a tiny charged particle. According to the Standard model of particle physics, it is a point particle. But the properties[7] of the electron gives the signature of some definite structure. From that point of view radius of electron is very significant and different electromagnetic phenomena revealed almost 8 different radii[4][6] of electron in different length scale. They are given here in Table-1.

\begin{table}
\caption{Eight different electron radii} 
\centering
\begin{tabular}{|c|c|}
\hline
Radius & Name\\
\hline
$R_0$ & Classical electron radius ($\frac{e^2}{mc^2}$) \\
\hline
$R_C$ & Compton radius ($\frac{\hbar}{mc}$) \\
\hline
$R_{QMC}$ & Quantum mechanical Compton radius ($\sqrt {3} \frac{\hbar}{mc}$) \\
\hline
$R_{QMC}^{\alpha}$ & QED-corrected quantum mechanical Compton radius ($\sqrt {3} (1+{\frac{\alpha}{2\pi}})\frac{\hbar}{mc}$) \\ 
\hline
$R_{em}$ & Classical electromagnetic radius ($\frac{{\hbar}^2}{me^2}$) \\
\hline
$R_H$ & Magnetic feild radius \\
\hline
$R_{QED}$ & Observed QED charge distribution for a bound electron\\
\hline
$R_E$ & Radius of the electric charge on the electron\\
\hline
\end{tabular}
\end{table}

\ As electron is a charged lepton with a small mass and definite magnetic moment, dipole moment, its behavior treated in terms of electrodynamics. First attempts were made by Lorentz and Abraham to give a structure of electron with the help of the dynamics of its charge[8]. Then quite a huge number of models are proposed by the scientists[8]. One recent and refined model is Relativistic Spinning Sphere (RSS) model, given by M. H. MacGregor[4]. 

\ Here in this article we aim to study the behavior of the self(internal) magnetic field produced by the rotation of charge with different radii of electron. The behavior of a rotating charge in uniform[2] and non-uniform external magnetic fields[3] are studied by Goldstein and Deissler. We study here the the magnetic field produced in the charged body itself for its own rotation as well as the current-loop calculations for different radii are also there.

\ We consider here a charged sphere which is rotating with the speed of light about its own axis of rotation which is in z-direction. Charge of the sphere is not glued over the sphere, rather it is contained in some very small space on the surface of the sphere. More clearly the glue over of the charge over the entire sphere of Compton sized model or equivalent is simply the violation of QED. Here our approach don't do that violation. Again the containing of charge is also gets experimental support. From recent LEP experiment it is concluded that the radius of electric charge on electron $R_E<10^{-19} m$. We suppose the charge of sphere is e, rest and mechanical mass is m and spin angular momentum is $\frac{\hbar}{2}.$ With the high speed rotation it can be assumed that mass of the body is at the equatorial plane without disrupting the original system.

\section{Rotation of charge and Ampere's law}
\paragraph \ 
When a charge rotates in a circular path, it creates a current loop. If the charge is e and the radius of the circular path is R, the current loop 
\begin{equation}{I = \frac{ev}{2\pi{R}}}\end{equation}
here v is the velocity with which the charge rotates about its axis.

In electrodynamics, the current I is also written in the form of current density J,
\begin{equation}{I = \int{J.da}}\end{equation} and $da$ is the area of the element. 
For magnetic field $B,$ we have
\begin{equation}{{\nabla \times B}={\mu}_0 J}\end{equation}
For S is the surface, C the curve and $dl$ is the line element, using Stoke's theorem,namely
\begin{equation}{{\int_{S}{(\nabla \times B)}.da}={\oint_{C}{B.dl}}}\end{equation} 
in (4), we have
\begin{equation}{{\oint_{C}{B.dl}}={{\frac{4\pi}{c}\int{J.da}}}={\mu}_0 I}\end{equation}
This is known as Ampere's law.

\ From equation (1) we have the expression for current loop with radius R when v is the velocity. As we have mentioned above that the electron has eight different radii from the view point of different electromagnetic phenomena, out of those we use here five radii to get different current contributions and magnetic field.

\ Compton radius of electron is given as $R_C = {\frac{\hbar}{mc}}.$  With the condition of the rotation of the charge with c velocity we get from equation (1)

\begin{equation}I_C={\frac{ec}{2\pi R_C}}= {\frac{c^2}{4\pi}}(\frac{em}{\frac{\hbar}{2}})\end{equation}

where e is the charge of electron, m is the mass of electron and $\frac{\hbar}{2}$ is the spin of the electron. 

\ So here we express the current loop in terms of three intrinsic properties of electron 
\begin{displaymath}I_C={\frac{ec}{2\pi R_C}}= {\frac{c^2}{4\pi}}(\frac{Charge.Mass}{Spin})\end{displaymath}

\ \ Classical electron radius is the first calculated classical radius and it was calculated in a non-relativistic way, $R_0={\frac{e^2}{mc^2}}$. In the similarly way like Compton radius for Classical electron radius we get the current loop as
\begin{equation}I_0={\frac{ec}{2\pi R_0}}= {\frac{c^2 {\alpha}^{-1}}{4\pi}}(\frac{em}{\frac{\hbar}{2}})\end{equation}

\ Classical electromagnetic radius is given by $R_{em}={\frac{{\hbar}^2}{me^2}}$ and for this radius the above expressions come out as

\begin{equation}I_{em}={\frac{ec}{2\pi R_{em}}}= {\frac{c^2 {\alpha}}{4\pi}}(\frac{em}{\frac{\hbar}{2}})\end{equation}

\ Quantum mechanical Compton radius comes from the concept of total angular momentum J and it is given as $R_{QMC}=\sqrt 3{\frac{\hbar}{mc}}$. The current-loop for Quantum mechanical Compton radius will be then

\begin{equation}I_{QMC}={\frac{ec}{2\pi R_{QMC}}}= {\frac{1}{\sqrt3}}{\frac{c^2}{4\pi}}(\frac{em}{\frac{\hbar}{2}})\end{equation}

\ QED-corrected quantum mechanical Compton radius includes the $(1+{\frac{\alpha}{2\pi}})$ term to the quantum mechanical Compton radius. The current-loop for QED-corrected Quantum mechanical Compton radius

\begin{equation}I^{\alpha} _{QMC}={\frac{ec}{2\pi R^{\alpha} _{QMC}}}= {\frac{(1-{\frac{\alpha}{2\pi}})}{\sqrt3}}{\frac{c^2}{4\pi}}(\frac{em}{\frac{\hbar}{2}})\end{equation}

\ Using the current contributions from different current-loop with five different radii we can get the Ampere's law. 

\ For $R_C$ the Ampere's law will be read as

\begin{equation}\oint_C B.dl = {\mu}_0 I= {\mu}_0 {\frac{c^2}{4\pi}}(\frac{em}{\frac{\hbar}{2}})\end{equation}

\ Similarly for  $R_0$, $R_{em}$, $R_{QMC}$ and $R^{\alpha} _{QMC}$ we have the Ampere's law 

\begin{equation}\oint_C B.dl = {\mu}_0 {\alpha}^{-1} {\frac{c^2}{4\pi}}(\frac{em}{\frac{\hbar}{2}})\end{equation}

\begin{equation}\oint_C B.dl = {\mu}_0 {\alpha}{\frac{c^2}{4\pi}}(\frac{em}{\frac{\hbar}{2}})\end{equation}

\begin{equation}\oint_C B.dl = {\frac{{\mu}_0}{\sqrt3}} {\frac{c^2}{4\pi}}(\frac{em}{\frac{\hbar}{2}})\end{equation}

\begin{equation}\oint_C B.dl = {\frac{{\mu}_0 (1-\frac{\alpha}{2\pi})}{\sqrt3}} {\frac{c^2}{4\pi}}(\frac{em}{\frac{\hbar}{2}})\end{equation}

respectively.

\ Using Schwinger correction[5][9] the magnetic moment of electron is expressed as $\mu \simeq {\frac{e\hbar}{2mc}}(1+{\frac{\alpha}{2\pi}}),$ with $\alpha$ = fine structure constant and this entirely is known as anomalous magnetic moment of electron. In other words, ${\mu}={\frac{e\hbar}{(m-\Delta m)c}}$ and $\Delta m \simeq m.{\frac{\alpha}{2\pi}}$ which gives ${\mu}\simeq{{\frac{e \hbar}{2mc}}(1+{\frac{\alpha}{2\pi}})}.$ The anomalous component is $\Delta m$ which is taken as the contribution of electromagnetic mass of the electron to the mechanical mass m. So the mass of electron we write here as a combination of mechanical mass and electromagnetic mass and mathematically we replace m by $m(1+{\frac{\alpha}{2\pi}})$.

\ Then all our above expression can be re-examined with the introduction of $m(1+{\frac{\alpha}{2\pi}})$ replacing mass m and re-written. 

\ Then the current-loop for Compton radius will be
 
\begin{displaymath}I_C = {\frac{c^2}{4\pi}}(\frac{em}{\frac{\hbar}{2}})(1+{\frac{\alpha}{2\pi}})\end{displaymath}

\ $(1+{\frac{\alpha}{2\pi}})$ can be taken as the first two terms of the series $e^{\frac{\alpha}{2\pi}}$ and hence we can write 

\begin{equation}I_C = {\frac{c^2}{4\pi}}(\frac{em}{\frac{\hbar}{2}}){e^{\frac{\alpha}{2\pi}}}\end{equation}

\ Similarly the current-loops for other four radii can be re-written as following

\begin{equation}I_0={\frac{ec}{2\pi R_0}}= {\frac{c^2 {\alpha}^{-1}}{4\pi}}(\frac{em}{\frac{\hbar}{2}}){e^{\frac{\alpha}{2\pi}}}\end{equation}

\begin{equation}I_{em}={\frac{ec}{2\pi R_{em}}}= {\frac{c^2 {\alpha}}{4\pi}}(\frac{em}{\frac{\hbar}{2}}){e^{\frac{\alpha}{2\pi}}}\end{equation}

\begin{equation}I_{QMC}={\frac{ec}{2\pi R_{QMC}}}= {\frac{1}{\sqrt3}}{\frac{c^2}{4\pi}}(\frac{em}{\frac{\hbar}{2}}){e^{\frac{\alpha}{2\pi}}}\end{equation}

\begin{equation}I^{\alpha} _{QMC}={\frac{ec}{2\pi R^{\alpha} _{QMC}}}= {\frac{1}{\sqrt3}}{\frac{c^2}{4\pi}}(\frac{em}{\frac{\hbar}{2}})\end{equation}

\ Hence in the same way the corrections we have to do in the expressions for Ampere's law written above.

\ New form for Compton radius will be as
\begin{equation}\oint_C B.dl = {\mu}_0 {\frac{c^2}{4\pi}}(\frac{em}{\frac{\hbar}{2}}){e^{\frac{\alpha}{2\pi}}}\end{equation}

\ Similarly for  $R_0$, $R_{em}$, $R_{QMC}$ and $R^{\alpha} _{QMC}$ we have the Ampere's law 

\begin{equation}\oint_C B.dl = {\mu}_0 {\alpha}^{-1} {\frac{c^2}{4\pi}}(\frac{em}{\frac{\hbar}{2}}){e^{\frac{\alpha}{2\pi}}}\end{equation}

\begin{equation}\oint_C B.dl = {\mu}_0 {\alpha}{\frac{c^2}{4\pi}}(\frac{em}{\frac{\hbar}{2}}){e^{\frac{\alpha}{2\pi}}}\end{equation}

\begin{equation}\oint_C B.dl = {\frac{{\mu}_0}{\sqrt3}} {\frac{c^2}{4\pi}}(\frac{em}{\frac{\hbar}{2}}){e^{\frac{\alpha}{2\pi}}}\end{equation}

\begin{equation}\oint_C B.dl = {\frac{{\mu}_0}{\sqrt3}} {\frac{c^2}{4\pi}}(\frac{em}{\frac{\hbar}{2}})\end{equation}

respectively.

\section{Approximation of long straight wire and B}
\paragraph \
To get the B, magnetic induction vector out of the integral above we use here the approximation of straight long current carrying wire[1] which results as $B={\frac{{\mu}_0 I}{2\pi R}}.$

So for $R_C$ \begin{displaymath}{B={\frac{{\mu}_0 I}{2\pi R_C}}}\end{displaymath}

Hence self magnetic field B for the electron rotating in a current loop with radius $R_C,$ the Compton radius comes out as \begin{equation}B_C = {\frac{{\mu}_0}{2\pi R_C}}{\frac{c^2}{4\pi}}(\frac{em}{\frac{\hbar}{2}}){e^{\frac{\alpha}{2\pi}}}\end{equation} 

\ Similarly we get for $R_{0},$ 

\begin{equation}B_0 = {\frac{{\mu}_0}{2\pi R_0}}{\frac{{\alpha}^{-1} c^2}{4\pi}}(\frac{em}{\frac{\hbar}{2}}){e^{\frac{\alpha}{2\pi}}}\end{equation} 

\ For $R_{em}$ we have

\begin{equation}B_{em} = {\frac{{\mu}_0}{2\pi R_{em}}}{\frac{{\alpha} c^2}{4\pi}}(\frac{em}{\frac{\hbar}{2}}){e^{\frac{\alpha}{2\pi}}}\end{equation}

\ For Quantum mechanical Compton radius $R_{QMC}$ B will be read as 

\begin{equation}B_{QMC} = {\frac{{\mu}_0}{2\pi R_{QMC}}}{\frac{c^2}{4\sqrt3\pi}}(\frac{em}{\frac{\hbar}{2}}){e^{\frac{\alpha}{2\pi}}}\end{equation}

and for QED-corrected Quantum mechanical Compton radius B will be

\begin{equation}B^{\alpha} _{QMC} = {\frac{{\mu}_0}{2\pi R^{\alpha} _{QMC}}}{\frac{c^2}{4\sqrt3\pi}}(\frac{em}{\frac{\hbar}{2}})\end{equation} 

\section{Self (internal) magnetic field H}
\paragraph \
Magnetic induction vector we get in the above section in terms of three intrinsic properties of electron. With using the well-known relation $H={\frac{1}{{\mu}_0}}B$ we get Self (internal) magnetic field for Compton radius

\begin{equation}H_C = {\frac{1}{{\mu}_0}}B_C = {\frac{1}{2\pi R_C}}{\frac{c^2}{4\pi}}(\frac{em}{\frac{\hbar}{2}}){e^{\frac{\alpha}{2\pi}}}\end{equation} 

\ With the Classical electron radius we have 

\begin{equation}H_0 = {\frac{1}{2\pi R_0}}{\frac{{\alpha}^{-1}  c^2}{4\pi}}(\frac{em}{\frac{\hbar}{2}}){e^{\frac{\alpha}{2\pi}}}\end{equation} 

\ Self(internal) magnetic field calculated for Classical electromagnetic radius is 
\begin{equation}H_{em} = {\frac{1}{2\pi R_{em}}}{\frac{{\alpha}c^2}{4\pi}}(\frac{em}{\frac{\hbar}{2}}){e^{\frac{\alpha}{2\pi}}}\end{equation} 

\ For Quantum mechanical Compton radius
\begin{equation}H_{QMC} = {\frac{1}{2\pi R_{QMC}}}{\frac{c^2}{4\sqrt3\pi}}(\frac{em}{\frac{\hbar}{2}}){e^{\frac{\alpha}{2\pi}}}\end{equation} 

and for QED-corrected Compton radius 

\begin{equation}H^{\alpha} _{QMC} = {\frac{1}{2\pi R^{\alpha} _{QMC}}}{\frac{c^2}{4\sqrt3\pi}}(\frac{em}{\frac{\hbar}{2}})\end{equation}

\section{Conclusion}
\paragraph \ 
The self-magnetic field of the electron is introduced here and it binds all three intrinsic properties of electron together in all of the above expressions for B and H. This immediately indicates the fact that for the self-magnetic field and accordingly for other related properties of electron, mass, charge and spin will have dominant roles. 

\ Both with mechanical and electromagnetic mass or anomalous magnetic moment contribution the expressions of B and H are found to be dependent on the mass, charge and spin together.

\ Expressions for B and subsequently H came out for all the five radii in a similar pattern with the inclusion of $(\frac{em}{\frac{\hbar}{2}})$ and $\frac{c^2}{4\pi}$ factor. Hence we put here $G=[\frac{c^2}{4\pi}(\frac{em}{\frac{\hbar}{2}})]$ as a constant for self-magnetic field and current-loop of electron. This constant G is in the expressions for current-loop and self-magnetic field we derived above and can be summerized in the forms shown below.

Hence the generalized expressions for current, magnetic induction vector and self-magnetic field are respectively

\begin{equation}I=G{e^{\frac{\alpha}{2\pi}}}\end{equation}

\begin{equation}B={\frac{{\mu}_0}{2\pi R}}G{e^{\frac{\alpha}{2\pi}}}\end{equation}

\begin{equation}H={\frac{G{e^{\frac{\alpha}{2\pi}}}}{2\pi R}}\end{equation}

\ Some numerical factors are there in expressions which come from the expressions of radii exclusively. In case of $R_0$ and $R_{em}$ the numerical factors are in the form of fine structure constant which is a dimensionless quantity. 

\ From the expressions of current-loop it comes out as a matter of fact that as smaller radius we treat larger current contributions will be in result. The relations between the different current-loops can be concluded as 

\begin{equation}I_C = {\alpha}{I_0}={\alpha}^{-1} {I_{em}} =\sqrt3{I_{QMC}} = \sqrt3 {e^{\frac{\alpha}{2\pi}}} {I^{\alpha} _{QMC}} \end{equation}
 
and similar relations we can get for H also

\begin{equation}H_C = {\alpha}^2 {H_0}={\alpha}^{-2} {H_{em}} =3{H_{QMC}} = 3 {e^{\frac{\alpha}{2\pi}}}^2 {H^{\alpha} _{QMC}} \end{equation}
  
\ Thus we have related current and magnetic field contributions for all the five radii mentioned above. The relations have shown here some $\alpha$-quantized results for Compton, Quantum mechanical Compton and QED-corrected quantum mechanical Compton radii.

\begin{figure*}
\centerline{
  \mbox{\includegraphics[width=4.0in]{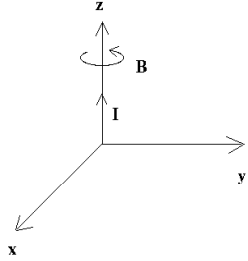}}
  }
    \caption{The co-ordinate system with the direction of current and magnetic field}
\label{overview}
\end{figure*}

\begin{figure*}
\centerline{
  \mbox{\includegraphics[width=4.0in]{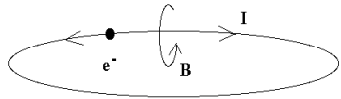}}
  }
    \caption{Current loop with and the direction of electron, current and magnetic field}
\label{overview}
\end{figure*}

\begin{figure*}
\centerline{
  \mbox{\includegraphics[width=4.0in]{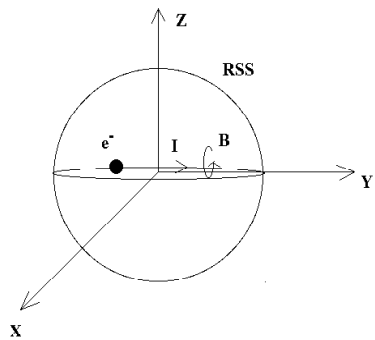}}
  }
    \caption{Current loop on the RSS model and self-magnetic field}
\label{overview}
\end{figure*}

\vspace{100pt}

\section{References}

\ \hspace{20pt} 1. D. J. Griffiths, Introduction to electrodynamics, 3rd edition,

\ \hspace{15pt} New delhi.

\ 2. R. J. Deissler, Phys. Rev. E 77, 036609 (2008)

\ 3. H. Goldstein, Am. J. Phys. 19, 100 (1951) 

\ 4. M. H. MacGregor, The Enigmatic Electron, Kluwer Academic
 
\ \hspace{15pt} Publishers, Dordrecht (1992).

\ 5. J. D. Jackson, Classical Electromagnetic Theory, Willey ,

\ \hspace{15pt}  Thidr Edition

\ 6. J-Marc, L-Leblond, Eur. J. Phys. 10 (1989) 265-268.

\ 7. Particle Data Group, “Review of Particle Properties”, 

\ \hspace{15pt} PRD 66 Part I, (2002)

\ 8. F. Rohrlich. Classical Charged Particles , Addison-Wesley, 

\ \hspace{15pt} Reading (1965)

\ 9. J. Schwinger, Phys. Rev. 73, 416 (1948)

\end{document}